\documentstyle[letter,epsf,wrapft,mbf]{ptptex}
\notypesetlogo  %comment in if to eliminate PTPTeX logo
\title{%        %You can use \\ for explicit line-break
Neutrino Interactions in Octet Baryon Matter
}
\author{%       %Use \sc for the family name
Toshitaka {\sc Tatsumi}$^{1,}$
\footnote{E-mail: tatsumi@ruby.scphys.kyoto-u.ac.jp} 
Tatsuyuki {\sc Takatsuka}$^{2,}$
\footnote{E-mail: takatuka@iwate-u.ac.jp} and 
Ryozo {\sc Tamagaki}$^{3,}$\footnote{E-mail: tamagaki@yukawa.kyoto-u.ac.jp} 
}

\inst{%         %Affiliation, neglected when [addenda] or [errata]
$^{1}$Department of Physics, Kyoto University, Kyoto 606-8502,Japan \\
$^{2}$Faculty of Humanities and Social Sciences, Iwate University \\
Morioka 020-8550, Japan \\
$^{3}$Kamitakano Maeda-cho 26-5, Kyoto 606-0097, Japan
}

\recdate{%      %Editorial Office will fill in this.
}

\abst{%       %this abstract is neglected when [addenda] or [errata]
Neutrino processes caused by the neutral current are studied in 
octet baryon matter. Previous confusion about the baryonic matrix 
elements of the neutral current interaction is excluded, and a correct 
table for them improved by consideration of the proton spin 
problem is presented instead.
}

\begin{document}

\maketitle

\newcommand{\gsim}{>\kern-12pt\lower5pt\hbox{$\displaystyle\sim$}}
\newcommand{\lsim}{<\kern-12pt\lower5pt\hbox{$\displaystyle\sim$}}
\newcommand{\Lam}{$\Lambda$}
\newcommand{\Sm}{$\Sigma$${}^{-}$}
\newcommand{\Cpc}{Cooper-pair cooling}
\newcommand{\beq}{\begin{equation}}
\newcommand{\eeq}{\end{equation}}
\newcommand{\beqa}{\begin{\eqnarray}}
\newcommand{\eeqa}{\end{\eqnarray}}

%\section{Introduction}
%Start your paper from here.

In neutron stars, neutrinos have played crucial roles and their weak 
interactions with nuclear matter have been intensively studied by 
many authors in relation to supernova explosions, thermal evolution 
of protoneutron stars and cooling of neutron stars.\cite{rf:1} 
There work two weak processes in the Standard model\cite{rf:2}; one is 
the charged current (CC) process and the other is the neutral current 
(NC) process. The elementary process where the CC takes part in has 
been well understood. On the other hand the NC process has not been 
well established yet due to little empirical information, while  
the NC process governs the neutrino mean-free path in baryonic matter 
and the $\nu\bar\nu$ emission from there. The former is closely related
to supernova explosions and deleptonization of protoneutron
stars.\cite{rf:1,rf:3}  The latter is related to 
the cooling problem of young 
neutron stars such as the 
$\nu\bar\nu$ bremsstrahlung process \cite{rf:4} and the $\nu\bar\nu$ 
emission due to the Cooper pairing of nucleons and quarks.\cite{rf:5,rf:6,rf:7}

In these contexts many authors have evaluated and used the matrix 
elements of the hadron NC between the baryon $SU(3)$ octet by utilizing 
$SU(3)$ symmetry. Nevertheless we have realized that there is 
some confusion spread over literatures about their values, especially 
for hyperons. The problem is related to the GIM mechanism.\cite{rf:2}  
There is another problem to be elucidated. 
It is well known that the same procedure has 
successively worked for the CC processes such as hyperon $\beta$ decays.
Although there has been no empirical justification a priori 
for the NC processes, recent experimental data and lattice simulations 
have suggested a large deviation from these values based on $SU(3)$ 
symmetry, which is caused by the existence of the flavor singlet 
current in the baryonic NC. The matrix elements of the flavor singlet 
current are closely related to the problem of the proton spin as well. 
To our knowledge there is no consideration to take into account these 
results for the NC processes in baryon matter. 

In this Letter we reconsider the NC processes in baryon matter to 
clarify the problematic points mentioned above and present
correct values about the matrix elements of the NC between the octet 
baryons. 

In some literatures studying neutron star matter with admixed hyperons, we notice such a mistaken statement that the 
NC $\Lambda\nu \bar{\nu}$ coupling vanishes, e.g., in Refs. 4b) and 6). 
As shown below, finiteness of this coupling is simply understood with 
use of the $SU(6)$ quark model for the baryon octet. The effective 
Lagrangian to describe the NC process is given by way of the 
Standard model,\cite{rf:2}
 \footnote{We suppress the $c,\;t,\;b$ quark sectors for our purpose.}
\begin{equation}
{\cal L}^{({\rm NC})}=\frac{G_{F}}{2\sqrt{2}}J_{\mu}j^{\mu},
\label{hamil}
\end{equation}
where $j^{\mu}$ is the lepton NC
, $j^{\mu}=\sum_{\ell =e,\mu,\tau}
\left[
\bar{\nu}_{\ell}\gamma^{\mu}(1-\gamma_{5})\;\nu_{\ell} 
-\bar{\ell}\gamma^{\mu}(1-4{\rm sin}^{2}\theta_{W}-\gamma_{5})\;\ell
\right],$
and $J_\mu$  is the quark NC current,
\begin{equation}
J_{\mu}=\bar{u}\gamma_{\mu}(I_{+}-\gamma _{5})u  
          -\bar{d}\gamma_{\mu}(I_{-}-\gamma _{5})d  
          -\bar{s}\gamma_{\mu}(I_{-}-\gamma _{5})s
\label{qnc}
\end{equation}
with $I_{+}=1-8\,{\rm sin}^{2}\theta _{W}/3$ and $I_{-}
=1-4\,{\rm sin}^{2}\theta _{W}/3$, where $\theta_{W}$ is the Weinberg 
angle (${\rm sin}^{2}\theta _{W}\simeq 0.23$). A point to be noted 
in what follows is the different sign between the terms 
of the upper components ($\nu_{\ell},\; u,\; c$) and the 
lower components ($\ell,\;d_c,\;s_c$) of the $SU(2)_{L}$ doublets. 
If we take the non-relativistic approximation to $J_{\mu}$, 
the vector (axial -vector) current operator becomes as 
$\bar{q}\gamma_{\mu}q\rightarrow \delta_{\mu ,0}q^{\dagger}q$ 
($\bar{q}\gamma_{\mu}\gamma_{5}q\rightarrow -\delta_{\mu ,k}
q^{\dagger}\sigma_{k}q$  with $k=1,2,3$). Then we can evaluate the matrix 
elements of $J_{\mu}$ between the baryon states described by 
the $SU(6)$ quark model wave functions, which 
are given as a sum of the 3-quark product wave functions. 
Because $J_{\mu}$ is the one-particle operator, only the diagonal 
(non-exchange) matrix elements with respect to the flavor and spin 
are non-vanishing. 

For the proton spin-up state described by
\[
|p \uparrow\rangle=\left[2(|u\uparrow u\uparrow d\downarrow\rangle+{\rm s.c.})
-(|u\uparrow u\downarrow d\uparrow\rangle+{\rm s.c.}) \right]/\sqrt{18},
\]
where $|u\uparrow u\uparrow d\downarrow\rangle$ etc. contain the 
antisymmetric color state and the symmetric spatial state and s.c. means the sum of symmetric combination, we have 
\begin{equation}
\langle p\uparrow|J_{\mu}|p\uparrow\rangle
=(2I_{+}-I_{-})\delta_{\mu,0}+(5/3)\delta_{\mu,3}
=(1-4\,{\rm sin}^{2}\theta_{W})\delta_{\mu,0}+(5/3)\delta_{\mu,3}.
\end{equation}
Similarly for the neutron state, because of $(I_{+}-2I_{-})=-1$, we have
\begin{equation}
\langle n\uparrow|J_{\mu}|n\uparrow\rangle
=-\delta_{\mu,0}-(5/3)\delta_{\mu,3}.
\end{equation}
The factor $5/3$ should be replaced by $g_{A}\simeq 1.27$.\cite{rf:8} 
%(the axial-vector to vector ratio of the coupling constant). 
For the $\Lambda$ state described by 
\[
|\Lambda \uparrow\rangle=\left[ |(u\uparrow d\downarrow-u\downarrow d\uparrow
-d\uparrow u\downarrow + d\downarrow u\uparrow)s\uparrow\rangle+{\rm s.c.}
\right] /\sqrt{12}, 
\]
we have 
\begin{equation}
\langle\Lambda\uparrow|J_{\mu}|\Lambda\uparrow\rangle
=-\delta_{\mu,0}-\delta_{\mu,3}.
\end{equation}
The non-vanishing $\Lambda\nu\bar{\nu}$ coupling is simply understood  
by the number counting, especially for the $\theta_{W}$-independent 
terms of the vector current by counting the number of each flavor quark 
with the positive (negative) sign for $u$ ($d,s$). The matrix elements 
for the other octet baryons are obtained in the same way.

We define the matrix elements of the baryonic NC as 
\beq
\langle B'|J_{\mu}|B\rangle\equiv \bar{u}'_{B}\gamma_{\mu}
(C_{V}^{SU(6)}-C_{A}^{SU(6)}\gamma_{5})u_{B},
\eeq
where $u_{B}$ ($u'_{B}$) is the Dirac spinor of the baryon in the 
initial (final) state. By equating the matrix elements obtained above 
with the non-relativistic limit of this equation, we can determine 
$C_{V, A}^{SU(6)}$ 
%the coefficients of the baryonic NC for the 
within the $SU(6)$ quark model: 
%Thus we have $C_{V}^{SU(6)}$ and $C_{A}^{SU(6)}$ as follows: 
$C_{V}^{SU(6)}=1-4\,{\rm sin}^{2}\theta_{W}$, 
$C_{A}^{SU(6)}=g_{A}$ for the proton, 
$C_{V}^{SU(6)}=-1$, $C_{A}^{SU(6)}=-g_{A}$ for the neutron, and 
$C_{V}^{SU(6)}=C_{A}^{SU(6)}=-1$ for the $\Lambda$ particle.  
In the same way we obtain the values for the other octet members.
$C_{V}^{SU(6)}$ thus obtained are the same as  
$C_{V}^{SU(3)}$ in Table I, while $C_{A}^{SU(6)}$ become the same 
as $C_{A}^{SU(3)}$ by taking $D=1$ and $F=2/3$ 
and replacing $D+F$ by $g_{A}$ in Table II. 

Next we study the properties of the matrix elements of the baryonic 
NC from a wider viewpoint, namely, on the basis of the SU(3)-symmetry 
consideration and with reference to the information obtained in 
 the proton spin problem. 

The quark NC current given in Eq. (2) is also written in 
the following form:
\begin{eqnarray}
J_\mu &=&(V^3_\mu-A^3_\mu)
-2\sin^2\theta_W\left(V^3_\mu+\frac{1}{\sqrt{3}}V^8_\mu\right)
+J^{\rm GIM}_\mu\nonumber\\
&\equiv&J^{\rm octet}_\mu+J^{\rm GIM}_\mu,
\label{hadron}
\end{eqnarray} 
where $V_\mu^\alpha, A_\mu^\alpha$
are the octet vector and axial-vector
currents defined by $V_\mu^\alpha={\bar q}\lambda^\alpha\gamma_\mu q$ 
and $A_\mu^\alpha={\bar q}\lambda^\alpha\gamma_\mu\gamma_5 q$ with 
the Gell-Mann matrices $\lambda^\alpha, ~\alpha=1\sim 8$. The current 
$J^{\rm octet}$ includes only the octet currents and $J^{\rm GIM}_\mu$ 
originates from the GIM mechanism and includes the $SU(3)$ singlet 
current; 
\beq
J^{\rm GIM}_\mu=\frac{1}{\sqrt{3}}
\left[V_\mu^8-\frac{1}{\sqrt{3}}V_\mu^0
-\left(A_\mu^8-\frac{1}{\sqrt{3}}A_\mu^0\right)\right],
\label{gim}
\eeq
where the $SU(3)$ singlet currents are defined by 
$V_\mu^0={\bar q}\lambda^0\gamma_\mu q$ and 
$A_\mu^0={\bar q}\lambda^0\gamma_\mu\gamma_5 q$ with 
$\lambda^0\equiv{\rm diag}(1,1,1)$. Note that $J^{\rm GIM}_\mu$ can be
explicitly written as $J^{\rm GIM}_\mu=-{\bar s}\gamma_\mu(1-\gamma_5)s$. 

The matrix elements of the octet currents between the baryon octet state
may be
easily evaluated by the Wigner-Eckert theorem and the Clebsch-Gordon
coefficients for $SU(3)$ symmetry.\cite{rf:2} We introduce the
coefficients $C_{V, A}^{\rm octet}$ for the octet
baryon state $|B\rangle$ by
\begin{eqnarray}
\lim_{q\to 0}\langle B'|J^{\rm octet}_\mu|B\rangle
 ={\bar u}_B'\gamma_\mu (C_V^{\rm octet}-C_A^{\rm octet}\gamma_5)u_B,
\label{octet}
\end{eqnarray}
as in Eq. (6). The coefficients $C_{V, A}^{\rm octet}$ then can be 
represented by the constants $D$ and $F$ \cite{rf:8,rf:9}, 
$D\simeq 0.80, F\simeq 0.47$. We list $C_{V, A}^{\rm octet}$ in Table I, II. 
The authors, e.g., in Refs. 4b) and 6) have used these values as 
the matrix elements of the baryonic weak NC and disregarded 
the contribution by $J^{\rm GIM}_\mu$. For $\Lambda$ and $\Sigma^{0}$ 
with $C_{V, A}^{\rm octet}=0$, the non-vanishing matrix elements 
come only from $J^{\rm GIM}_\mu$ written by the $s$-quark operator. 
%%GIM mechanism
%%Wrong for Maxwell, Yakovlev
To evaluate the matrix elements of $J^{\rm GIM}_\mu$ we need a special
care, because it includes the $SU(3)$ singlet currents, especially the 
$U_A(1)$ current; since the singlet vector current $V_\mu^0$ is
nothing but three times the baryon $U_{V}(1)$ current and thereby
conserved, the matrix element should be easily evaluated, 
$\langle B'|V_\mu^0|B\rangle=3{\bar u}_B'\gamma_\mu u_B$. 
A naive thinking that $J^{\rm GIM}_\mu$ gives no contribution for 
nucleons since it is written only in terms of strange quark
%, $J_\mu^{\rm GIM}=-{\bar s}\gamma_\mu(1-\gamma_5)s$, 
gives a value of $C_A^0$ defined by
\beq
\langle B'|A_\mu^0|B\rangle=C_A^0{\bar u}'_B\gamma_\mu\gamma_5 u_B;
\label{ua1}
\eeq
e.g. $C_A^0({SU(6)})=1$ within the non-relativistic $SU(6)$ quark model
and $C_A^0({SU(3)})=-(D-3F)\simeq 0.61$ by way of $SU(3)$
symmetry. Thus we can give the values of $C_{V, A}^{\rm GIM}$ 
defined like in Eq.~(\ref{octet}). The sum of $C_{V, A}^{\rm octet}$ 
and $C_{V, A}^{\rm GIM}$,
$C_{V, A}=C_{V, A}^{\rm octet}+C_{V, A}^{\rm GIM}$ then gives 
the baryon matrix elements. We present these $C_{V, A}$ 
in Table I, II within $SU(3)$ symmetry, denoted by 
$C_{V,A}^{SU(3)}$. Some recent literatures have used these expressions, e.g. Refs. 3) and 7).

\begin{table}
\begin{center}
\caption{Vector coefficients $C_V$.}
\begin{tabular}{|c|c|c|}
\hline
 $B$ &  $C_V^{\rm octet}$   & $C_{V}^{SU(3)}=C_V$ 
  \\ \hline 
  $n$ & $-1$ & $-1$    \\ 
$p$ & ~$1-4\,{\rm sin}^{2}\theta_{W}$~ &
$~1-4\,{\rm sin}^{2}\theta_{W}$~   \\ 
$\Lambda$ & 0 & $-1$  \\ 
$\Sigma^{-}$ & 
$ ~-(2-4\,{\rm sin}^{2}\theta_{W})$~ & ~$-(3-4\,{\rm sin}^{2}\theta_{W})$~ \\ 
$\Sigma^{+}$ & 
$2-4\,{\rm sin}^{2}\theta_{W}$ & $1-4\,{\rm sin}^{2}\theta_{W}$ \\ 
$\Sigma^{0}$ & 0 & $-1$  \\ 
$\Xi^{-}$ & $-(1-4\,{\rm sin}^{2}\theta_{W})$ & $-(3-4\,{\rm sin}^{2}
\theta_{W})$\\ 
$\Xi^{0}$ & 1 & $-1$
 \\ \hline
\end{tabular}
\end{center}
\end{table}

\begin{table}
\begin{center}
\caption{Axial-vector coefficients $C_A$.}
\begin{tabular}{|c|c|c|c|}
\hline
 $B$ &  $C_A^{\rm octet}$   & $C_{A}^{SU(3)}$ & $C_A$  \\ 
\hline 
  $n$ & ~$-(D+F)$~ & $-(D+F)$ &  $-4/3D- \Sigma/3$ \\ 
$p$ & ~$D+F$~ & ~$D+F$~ & $2/3D +2F- \Sigma/3$ \\ 
$\Lambda$ & 0 & ~$-(D/3+F)$~ & $-2/3D- \Sigma/3$ \\ 
$\Sigma^{-}$ & $-2F$ & $D-3F$  & $2/3D-2F- \Sigma/3$\\ 
$\Sigma^{+}$ & $2F$ & $D+F$ & ~$2/3D+2F- \Sigma/3$~  \\ 
$\Sigma^{0}$ & $0$ & $D-F$ & $2/3D- \Sigma/3$  \\ 
$\Xi^{-}$ & $D-F$ & $D-3F$ & $2/3D-2F- \Sigma/3$\\ 
$\Xi^{0}$ & $-(D-F)$ & $-(D+F)$ & $-4/3D- \Sigma/3$
 \\ \hline
\end{tabular}
\end{center}
\end{table}

In 1988 EMC presented information about the baryon matrix element
of the singlet current. They measured so-called the quark content of
the proton, which triggered the exploding studies about the proton spin problem. Take a proton with momentum $p$ and spin $s$ and consider the following matrix element of the $U_A(1)$ current measured at the 
scale $Q^2$,\cite{rf:9}
\beq
\Sigma(Q^2)s_\mu=\Biggl\langle p,s\left|\sum_{u,d,s}{\bar q}_i
\gamma_\mu\gamma_5q_i\left|_{Q^2}\right.\right|p,s\Biggr\rangle.
\label{spin}
\eeq
Writing $\Sigma(Q^2)\equiv \Delta u+\Delta d+\Delta s$, $\Delta q$ is
called the quark content inside the proton. Then the coefficient 
$C_A^0$ can be identified with $\Sigma(Q^2\sim 0)$ at low 
momentum transfer. EMC measured the spin-dependent proton structure function 
by using the polarized muon beam and proton at 
$Q^2_{\rm EMC}=10.7{\rm GeV}^2$,\cite{rf:10} and extracted 
a remarkably small value. 
Subsequently SMC also reported a small value at 
$Q^2_{\rm SMC}=5{\rm GeV}^2$ \cite{rf:11} and confirmed the
EMC conclusion. Although there are still left some ambiguities about 
the value, they suggest 
$C_A^0=0 - 0.32$. \cite{rf:12}
%\Sigma(Q_{\rm EMC}^2)=0.13\pm 0.19.
%\Sigma(Q_{\rm SMC}^2)=0.23\pm 0.06.
This value is quite discrepant with the $SU(3)$ value. 
Subsequent lattice simulations have also shown 
$C_A^0=0.08 - 0.37$,
which is consistent with the observations.\cite{rf:13}

\begin{wrapfigure}{r}{6.6cm}
  \epsfxsize=6cm 
  \epsffile{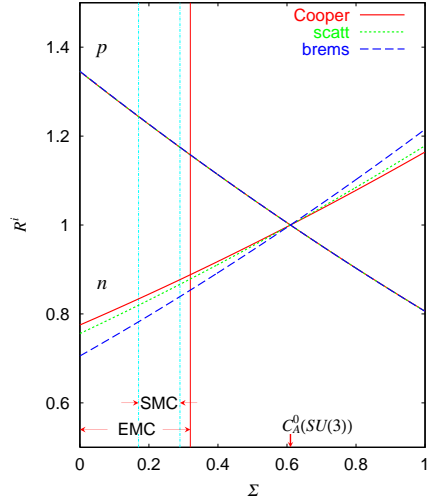}
  \caption{Ratios of the reaction rates as functions of $\Sigma$. Upper
 three curves show the ratios for the proton processes, which are almost
 overlapped with each other due to the fact, $C_V\sim 0$. Lower three curves show the ratios for the
 neutron processes.}
 \label{fig:1}
\end{wrapfigure}

%\begin{wrapfigure}{r}{6.6cm}
%  \epsfxsize=6cm 
%  \epsffile{ptpnifig1.eps}
 % \caption{Change (versus $\Sigma$) of the neutrino-proton scattering 
%cross section $\sigma _{\nu p}$ and the neutrino emissibity of 
%Cooper-pair cooling due to the neutron ${}^3 P_{2}$ superfluid, 
%relative to the values given in $SU(3)$ symmetry.}
% \label{fig:1}
%\end{wrapfigure}

The $U_A(1)$ current is not conserved even when quarks are massless, 
and has an anomalous divergence coming from the triangle diagram,
\beq
\partial^{\mu} A_\mu^0=-\frac{3\alpha_s}{2\pi}{\rm Tr}G\tilde G,
\eeq
where $G_{\mu\nu}$ is the gluon tensor (the Adler-Bell-Jakiw
anomaly). Hence $\Sigma(Q^2)$ should consist of not only the quark
but also the gluon contributions, the latter of which can be 
also interpreted as the sea-quark contribution. 
Thus we can write $\Sigma(Q^2)$ as\cite{rf:14,rf:9} 
\beq
\Sigma(Q^2)=\Sigma_{\rm quark}-\frac{3\alpha_s(Q^2)}{2\pi}\Delta g(Q^2),
\label{tria}
\eeq
where $\Delta g(Q^2)$ is an integrated gluon 
distribution inside the proton. Then $\Sigma_{\rm quark}$ can be identified 
with the $C_A^0({SU(3)})$ for the proton. Extending this idea to other
members by way of $SU(3)$ symmetry, we can evaluate
$C_A$ for the baryon octet by taking into account the sea-quark contribution.
In Table I, II we list the final expressions of $C_{V, A}$ 
corrected by these 
considerations, i.e., $C_{A}=C_{A}^{SU(3)}-(\Sigma-C_A^0({SU(3)}))/3
=C_{A}^{SU(3)}-D/3+F-\Sigma/3$.

To demonstrate the effect of $\Sigma$ on some neutrino processes in
neutron stars, we consider the $\nu\bar\nu$ bremsstrahlung emissivity 
$\propto C_A^2$, the 
$\nu\bar\nu$ emissivity due to the $^3P_2$-Cooper pairing $\propto 
C_V^2+2C_A^2$ and
the $\nu(\bar\nu)$ scattering opacity $\propto C_V^2+3C_A^2$ 
in nuclear matter. 
%The ratios of
%the reaction rates to the $SU(3)$ values then read
%$R^{\rm brems}=(C_A/C_A^{SU(3)})^2$ \cite{rf:4}, 
%$R^{\rm Cooper}=(C_V^2+2C_A^2)/(C_V^{SU(3)2}+2C_A^{SU(3)2})$ \cite{rf:6}
%and $R^{\rm scatt}=(C_V^2+3C_A^2)/(C_V^{SU(3)2}+3C_A^{SU(3)2})$ \cite{rf:1}, 
%respectively. 
In Fig.~1 we plot their ratios $R^i$ to the $SU(3)$ values 
as functions of $\Sigma$.
We can easily see that $R^i$ are given as increasing functions
for neutrons, while decreasing functions for protons; it also implies that 
the sea-quark contribution suppresses the neutron processes at most
by 30\%, while 
enhances the proton processes at most by 40\%. 

%Effects caused by the change of $C_{A}$ from $C_{A}^{SU(3)}$ 
%are demonstrated for two examples relevant to nucleons: 
%One is the total neutrino-proton scattering cross section 
%$\sigma_{\nu p}\propto (C_{V}(p)^{2}+3C_{A}(p)^{2})$\cite{rf:1}
%and the other is the neutrino emissibity 
%$\epsilon_{\nu} \propto (C_{V}(n)^{2}+2C_{A}(n)^{2})$ of the 
%Cooper-pair cooling of neutron stars due to the neutron ${}^3 P_{2}$ 
%superfluid.\cite{rf:6} In Fig. 1 we show the ratio of $\sigma_{\nu p}$ 
%and $\epsilon_{\nu} $ to those given by the $SU(3)$ value. 
% We see a noticeable change as $\Sigma$ changes from 
%$C_{A}^{SU(3)}=0.675$ to zero; $\sigma_{\nu p}$ increases 
%by about 40\% and $\epsilon_{\nu}$ decreases by about 25\%.

We have assumed $SU(3)$ symmetry for baryons and the sea-quark 
contribution as well, but it is not obvious, especially for the 
see-quark contribution. When one uses the $SU(3)$ value for $C_A^0$ 
and the experimental data of $\Sigma(Q^2)$, one may extract 
the sea quark contribution to be, e.g. ,  
$\Delta_{\rm sea}u \simeq\Delta_{\rm sea}d\simeq
\Delta_{\rm sea}s \simeq-0.16$ from the EMC data.\cite{rf:10}  
Thus the sea-quark contribution is almost $SU(3)$ symmetric.
Recent lattice simulations have studied the $SU(3)$ symmetry breaking
effect, and also supported our idea that the sea-quark contribution is 
approximately $SU(3)$ symmetric.\cite{rf:13} 

Although we can estimate the value of $C_A^0$ by using the data of
deep inelastic scattering, a direct measurement of $C_A^0
=\Sigma(Q^{2}\sim 0)$ may be carried out by the $\nu, \bar{\nu}$ 
elastic scattering off nucleons or nuclei.\cite{rf:15}

\section*{Acknowledgements}
We thank T. Muto, S. Tsuruta and H. Matsufuru for useful discussions. 
The present work is supported by the 
Japanese Grant-in-Aid for Scientific Research Fund of the Ministry
of Education, Culture, Sports, Science and Technology (11640272,
13640282, 13640252).

\end{document}